\documentstyle[osa,aps,prl,epsfig]{revtex}
\begin{document}
\twocolumn[\hsize\textwidth\columnwidth\hsize\csname
@twocolumnfalse\endcsname
\title{%$^{\ast\ast\ast}$
Hole-driven MIT theory, Mott transition in VO$_2$,
MoBRiK$^{\ast}$}
\author{$^{\dagger}$Hyun-Tak Kim, Bong-Jun Kim, Yong Wook Lee, Byung-Gyu Chae, Sun Jin Yun, and Kwang-Yong Kang}
\address{IT Convergence and Components Research Lab., ETRI, Daejeon 305-350, Republic of
Korea}
\date{July 22, 2006}
\maketitle{}
%\newpage
\begin{abstract}
For inhomogeneous high-$T_c$ superconductors, hole-driven
metal-insulator transition (MIT) theory explains that the gradual
increase of conductivity with increasing hole doping is due to
inhomogeneity with the local Mott system undergoing the
first-order MIT and the local non-Mott system. For VO$_2$, a
monoclinic and correlated metal (MCM) phase showing the linear
characteristic as evidence of the Mott MIT is newly observed by
applying electric field and temperature. The structural phase
transition occurs between MCM and Rutile
metal phases. Devices using the MIT are named MoBRiK.\\ \\

%PACS numbers: 71.27. +a, 71.30.+h
\end{abstract}
]
%\newpage
 It has been generally accepted that conductivity,$\sigma$,  and $T_c$ for
 inhomogeneous high-$T_c$ superconductors [1] as strongly correlated
 systems gradually increase with doped hole density in a Mott insulator
 from under-doping to critical doping [2], although a first-order
 transition near the Mott insulator had been theoretically suggested [3].
 These phenomena seemed to be explained by the Mott-Hubbard (MH)
 continuous metal-insulator transition (MIT) theory. However, since
 the MH theory was established for homogeneous system, the theory
 does not explain the phenomena in inhomogeneous system. The first-order Mott
 MIT without the structural phase transition (SPT) has not clearly been proved.

This paper briefly describes important ideas and physical meanings
of the hole-driven MIT theory  (extended Brinkman-Rice (BR)
picture [4]) named by a reviewer in ref. 7 and explains the above
phenomena. An experimental observation is also presented to
clarify the Mott MIT.

Metal has the electronic structure of one electron per atom in
real space, i.e. half filling, which indicates that $\delta$q=
$\sum$(q$_i$-q$_j$)=0 where q$_i$ and q$_j$ are charge quantities
at $i$ and $j$ nearest neighbor sites, respectively [4]; there is
no charge density wave. The BR picture explains physical
properties of a strongly correlated metal, which was developed
when $n=l$ for a homogeneous system with one electron per atom,
where $n$ is  the number of electrons in the Mott system with the
electronic structure of metal (Fig. 1(a) left) and $l$ is the
number of lattices in the measurement region [3]. The Mott
insulator becomes metal via MIT.

When n$<$l (inhomogeneous) (Fig. 1(a) right), Fourier transform
from K-space to real-space is not satisfied. The local Mott
insulator (system) becomes metal after MIT. When the inhomogeneous
system is measured, local metal (Mott) regions are averaged over
all lattice sites (measurement region); the measured physical
quantity is an averaged one. Then the effective fractional charge
is given by $e^{\prime}={\rho}e$, where 0$<\rho=n/l\le$1 is band
filling. The fractional Coulomb energy between quasiparticles is
defined by $U={\rho}^2U^{\prime}=\kappa\rho^2U_c$, where
0$<\kappa<$1 ($\kappa=U^{\prime}/U_c$ is the correlation
strength), 0$<\kappa\rho^2<$1, $U_c$ is the critical on-site
Coulomb interaction in the BR picture [3]. The system is satisfied
with Fourier transform. The physical meaning of the fractional
$e^{\prime}$ and $U$ is the effect of measurement (average) and
not true effect.

The averaged system has an electronic structure of one effective
charge per atom; $\rho^{\prime}=n^{\prime}/l$=1 where $n^{\prime}$
is the number of the effective charges. This is the same
electronic structure as one of the BR picture with $\kappa$=1
[3,4]. The effective mass of the quasiparticle is derived from
Gutzwiller's variational method and has the same form as that in
the BR picture. This is because the averaged system with
$\rho^{\prime}$=1 and the true system satisfied with the BR
picture are mathematically self-consistent. The effective mass is
given by

\begin{eqnarray}
\frac{m^*}{m}&=&\frac{1}{1-(\frac{U}{U_c})^2}=\frac{1}{1-{\kappa}^2{\rho}^4},
~~~0<\kappa\rho^2<1. \nonumber
\end{eqnarray}

\noindent In the experimental analysis for an inhomogeneous
system, $\kappa$ was estimated as one [4]. The observed effective
mass is given by
\begin{eqnarray}
\frac{m^*}{m}&=&\frac{1}{1-{\rho}^4}.
\end{eqnarray}

\noindent Eq. (1) is defined at $\rho\ne$1 (Fig. 1(b)) and has a
first-order discontinuous MIT between a Mott insulator with $U_c$
at $\rho$=1 and a metal at $\rho_{max}<$1. The MIT is caused due
to breakdown of $U_c$ by hole doping of a very low density,
$n_c=1-\rho_{max}$, into the Mott insulator (Fig. 1(c)); this is a
hole-driven MIT. $n_c$ is a minimum constant hole density where
the MIT occurs. Conversely, control of nc makes the Mott insulator
switch between insulator and metal. After the MIT, the local Mott
insulators become strongly correlated local metals with a
$\kappa\ne$1 value in the BR picture. $m^{\ast}$ in Eq. (1) is an
average of the true effective mass in the BR picture.

For inhomogeneous superconductors, the gradual increase of
conductivity with doping [2] is that the local Mott insulators in
(Fig. 1(a) right) continuously change into metal with hole doping;
$\sigma\propto(m^{\ast}/m)^2$ with doping density, $\rho$,
dependence in Eq. (1). The reason why the measured conductivity
with doping is continuous is that the magnitude of a local metal
conductivity after the MIT is very small because the local area is
within about 3nm [1].
\begin{figure}
\vspace{-0.8cm}
\centerline{\epsfysize=14.0cm\epsfxsize=8.8cm\epsfbox{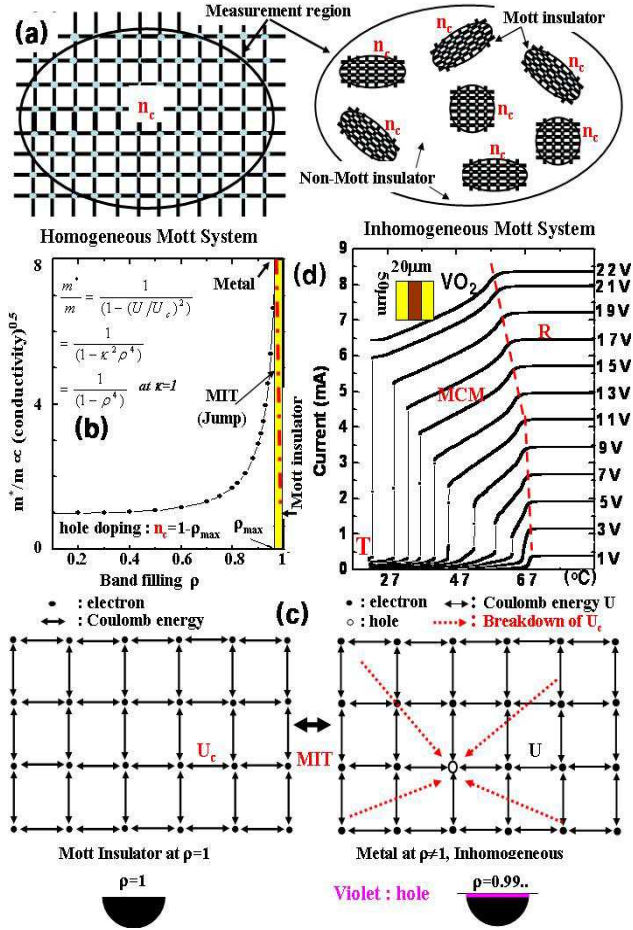}}
\vspace{-0.0cm} \caption{(a) Left: homogeneous Mott system. Right:
inhomogeneous system with local Mott insulators where the
first-order MIT occurs and non-Mott insulator regions where the
MIT does not occur. Example: Mott insulator of VO$_2$, non-Mott
insulator of V$_2$O$_5$. (b) The effective mass of Eq. 1 with
divergence near band filling $\rho$=1. Red-dash line indicates the
first-order MIT. Yellow region is critical hole density $n_c$ in
which MIT occurs.  (c) The physical meaning of Eq. (1). Breakdown
of $U_c$ is shown by hole doping of a low density into Mott
insulator; this is the MIT. (d) Electric field and temperature
dependence of the MIT measured for a VO$_2$-based two terminal
device (Inset, brown: VO$_2$; yellow: electrodes).  The MCM metal
phase showing the linear characteristic is clearly shown.}
\end{figure}
Furthermore, the Mott MIT [5-7] and the Peierls MIT with the SPT
[8,9] have been controversial even in VO$_2$ as a representative
Mott insulator. This is due to ambiguity of relation between MIT
and SPT. Fig. 1(d) clearly shows the relation. VO$_2$ was known to
simultaneously undergo the MIT and the SPT near 68$^{\circ}$C
(similar to V=1 case in Fig. 1(d)). Actually, even in this case,
the MIT was not simultaneous with the SPT in our work [7]. When
voltage is applied to a VO$_2$-based device [7], the MIT occurs
between T (monoclinic semiconductor phase, transient triclinic)
[5,6] and MCM (monoclinic and  correlated metal) phase with the
increase of $\sigma$. The SPT happens between MCM and R (rutile
tetragonal metal phase) (SPT instability, red-dash line in Fig.
1(d)); this was confirmed by micro-Raman experiment [10]. This
indicates that the MIT is not related to the SPT. The MCM phase as
evidence of the Mott transition clarifies the controversial
problem. MCM differs from Pouget et al.'s M$_2$ Mott-Hubbard
insulator phase [5,6]. We consider that T can be the paramagnetic
Mott insulator with the equally spaced chain structure because T
and MCM have the same structure.

MCM is caused by $n(E)=n_c(T,E)-n(T)$, where $n(E)$ is the hole
density excited by electric field (voltage), $n(T)$ is the hole
density excited by temperature, $n_c(T,E)$ is the critical hole
density in which the MIT occurs by the electric field and
temperature excitations [7]. Hole carriers were confirmed by Hall
measurement [7]. For constant $n_c$, $n(T)$ decreases as $n(E)$
increases. Thus, $T_{MIT}$ decreases, which is evidence of the
fact that the MIT is controlled by doped holes. This is predicted
in Eq. (1). Further, MCM is regarded as a non-equilibrium state
because metal exists at the divergence in Eq. (1) (Fig. 2(b)).
First-order MIT devices using the transition between T and MCM are
called $\bf{MoBRiK}$ from names of
$\bf{Mo}$tt-$\bf{B}$rinkman-$\bf{Ri}$ce-$\bf{K}$im physicists who
have established and extended the MIT.

%\narrowtext
\end{document}